\documentclass[10pt,journal,compsoc]{IEEEtran}
\usepackage[normalem]{ulem}
\ifCLASSOPTIONcompsoc
  \usepackage[nocompress]{cite}
\else
  \usepackage{cite}
\fi
\ifCLASSINFOpdf
  \usepackage[pdftex]{graphicx}
  \DeclareGraphicsExtensions{.pdf,.jpeg,.png}
\else
\fi
\usepackage{amsmath}
\usepackage[scientific-notation=true]{siunitx}
\usepackage{amsfonts}
\usepackage{soul}
\usepackage{tikz}
\usetikzlibrary{matrix}
\usepackage{algorithm}
\usepackage{algorithmicx}
\usepackage{algpseudocode}

\usepackage{multirow}
\usepackage{array}
\usepackage{array}
  \usepackage[caption=false,font=footnotesize]{subfig}
\usepackage{hyperref}
\usepackage[T1]{fontenc}
\usepackage{color}
\usepackage{booktabs}

\algrenewcommand\algorithmicrequire{\textbf{Input:}}
\algrenewcommand\algorithmicensure{\textbf{Output:}}
\DeclareMathOperator{\Log}{log}

\begin{document}
%
% paper title
% Titles are generally capitalized except for words such as a, an, and, as,
% at, but, by, for, in, nor, of, on, or, the, to and up, which are usually
% not capitalized unless they are the first or last word of the title.
% Linebreaks \\ can be used within to get better formatting as desired.
% Do not put math or special symbols in the title.
\title{QR factorization of ill-conditioned tall-and-skinny matrices on distributed-memory systems}
%
%
% author names and IEEE memberships
% note positions of commas and nonbreaking spaces ( ~ ) LaTeX will not break
% a structure at a ~ so this keeps an author's name from being broken across
% two lines.
% use \thanks{} to gain access to the first footnote area
% a separate \thanks must be used for each paragraph as LaTeX2e's \thanks
% was not built to handle multiple paragraphs
%
%
%\IEEEcompsocitemizethanks is a special \thanks that produces the bulleted
% lists the Computer Society journals use for "first footnote" author
% affiliations. Use \IEEEcompsocthanksitem which works much like \item
% for each affiliation group. When not in compsoc mode,
% \IEEEcompsocitemizethanks becomes like \thanks and
% \IEEEcompsocthanksitem becomes a line break with idention. This
% facilitates dual compilation, although admittedly the differences in the
% desired content of \author between the different types of papers makes a
% one-size-fits-all approach a daunting prospect. For instance, compsoc 
% journal papers have the author affiliations above the "Manuscript
% received ..."  text while in non-compsoc journals this is reversed. Sigh.

\author{Nenad~Miji\' c,
        Abhiram~Kaushik,
        and~Davor~Davidovi\' c% <-this % stops a space
\IEEEcompsocitemizethanks{\IEEEcompsocthanksitem N. Miji\' c, A. Kaushik and D. Davidovi\' c was with the Centre for Informatics and Computing, Ru\dj er Bo\v skovi\'c Institute, Zagreb,
Croatia.\protect\\
% note need leading \protect in front of \\ to get a newline within \thanks as
% \\ is fragile and will error, could use \hfil\break instead.
E-mail: nmijic@irb.hr, abadrin@irb.hr, ddavid@irb.hr.}% <-this % stops an unwanted space
%\thanks{Manuscript received ???, 2024; revised ???, 2024.}
}

\IEEEtitleabstractindextext{%
\begin{abstract}

In this paper we present a novel algorithm developed for computing the QR factorisation of extremely ill-conditioned tall-and-skinny matrices on distributed memory systems.
The algorithm is based on the communication-avoiding CholeskyQR2 algorithm and its block Gram-Schmidt variant. The latter improves the numerical stability of the CholeskyQR2 algorithm and significantly reduces the loss of orthogonality even for matrices with condition numbers up to $10^{15}$. Currently, there is no distributed GPU version of this algorithm available in the literature which prevents the application of this method to very large matrices. In our work we provide a distributed implementation of this algorithm and also introduce a modified version that improves the performance, especially in the case of extremely ill-conditioned matrices.
The main innovation of our approach lies in the interleaving of the CholeskyQR steps with the Gram-Schmidt orthogonalisation, which ensures that update steps are performed with fully orthogonalised panels.
The obtained orthogonality and numerical stability of our modified algorithm is equivalent to CholeskyQR2 with Gram-Schmidt and other state-of-the-art methods.
Weak scaling tests performed with our test matrices show significant performance improvements. In particular, our algorithm outperforms state-of-the-art Householder-based QR factorisation algorithms available in ScaLAPACK by a factor of $6$ on CPU-only systems and up to $80\times$ on GPU-based systems with distributed memory.

\end{abstract}

% Note that keywords are not normally used for peerreview papers.
}

% make the title area
\maketitle

% To allow for easy dual compilation without having to reenter the
% abstract/keywords data, the \IEEEtitleabstractindextext text will
% not be used in maketitle, but will appear (i.e., to be "transported")
% here as \IEEEdisplaynontitleabstractindextext when the compsoc 
% or transmag modes are not selected <OR> if conference mode is selected 
% - because all conference papers position the abstract like regular
% papers do.
\IEEEdisplaynontitleabstractindextext
% \IEEEdisplaynontitleabstractindextext has no effect when using
% compsoc or transmag under a non-conference mode.

% For peer review papers, you can put extra information on the cover
% page as needed:
% \ifCLASSOPTIONpeerreview
% \begin{center} \bfseries EDICS Category: 3-BBND \end{center}
% \fi
%
% For peerreview papers, this IEEEtran command inserts a page break and
% creates the second title. It will be ignored for other modes.
\IEEEpeerreviewmaketitle

%\IEEEraisesectionheading{\section{Introduction}\label{sec:introduction}}
% Computer Society journal (but not conference!) papers do something unusual
% with the very first section heading (almost always called "Introduction").
% They place it ABOVE the main text! IEEEtran.cls does not automatically do
% this for you, but you can achieve this effect with the provided
% \IEEEraisesectionheading{} command. Note the need to keep any \label that
% is to refer to the section immediately after \section in the above as
% \IEEEraisesectionheading puts \section within a raised box.

\IEEEraisesectionheading{\section{Introduction}\label{sec:intro}}

\IEEEPARstart{I}{n} this paper, we develop and analyse scalable algorithms tailored to large-scale distributed-memory systems that can compute the
QR decomposition~\cite{Golub2013MatrixComputations} of a rectangular matrix $A \in \mathbb{R}^{m\times n}$ with $m \geq n$:

\begin{equation}
    A = Q R,
\end{equation}

\noindent 
where $Q \in R^{m\times n}$ is an orthogonal matrix and $R\in\mathbb{R}^{n\times n}$ is an upper triangular matrix. Finding the QR decomposition of large rectangular matrices is a crucial step in various numerical methods such as block methods (e.g. in the solution of linear systems with multiple right-hand sides \cite{Heyouni2005MatrixSides}), the Krylov subspace methods \cite{Gutknecht2006BlockIntroduction}, eigenvalue solvers (e.g. in the reduction to the band form of the multi-stage eigensolver \cite{Davidovic2012ApplyingGPUs}), subspace iteration \cite{Winkelmann2019} and solving a dense least squares problem of an overdetermined system. In some extreme cases, where $m \gg n$, i.e. the matrix has many more rows than columns (so-called tall and skinny matrix - TS), the calculation of the QR factorisation becomes a critical path and requires special methods. One example is a subspace projection iterative eigensolver~\cite{Winkelmann2019} to calculate a small fraction of the extreme eigenvalues.

The most stable and accurate solution to compute QR factorisation for general matrices is based on Householder reflectors \cite{Golub2013MatrixComputations}, usually referred to as Householder QR. The algorithm finds a set of orthogonal Householder matrices that annihilate the entries below the main diagonal of $A$ column by column, while the rest of the matrix is updated (trailing update). This type of algorithm, where $R$ is constructed by applying an orthogonal matrix from the left, $Q^T A = R$, is called orthogonal triangularisation and provides good numerical stability. The highly efficient implementations can be found in many numerical libraries such as ScaLAPACK \cite{Choi1996b} for distributed memory systems or MAGMA \cite{Haidar2015MAGMAComputing} for heterogeneous and accelerator-based shared memory systems. However, the Householder QR cannot achieve high performance for tall and skinny matrices~\cite{Demmel2012Communication-optimalFactorizations} because the panel factorisation of columns with many more rows is performed with much slower Level-1 and Level-2 BLAS kernels, which cannot be compensated by the highly parallel and optimised Level-3 BLAS kernels used in the trailing update \cite{blackford2002updated}.

To avoid working with very tall and skinny panels, a Tall Skinny QR (TSQR) algorithm \cite{Demmel2012Communication-optimalFactorizations, Demmel2008Communication-avoidingFactorizations} was developed, which offers better parallelisation on systems with distributed memory and reduces communication. The main idea is to split the input matrix into a cyclic 1-D block row layout so that QR factorisations can be performed on local blocks concurrently. In the following reduction steps, the intermediate R-factors are grouped into pairs and orthogonalised. This step is repeated until the final upper triangular $R$ is reached. Although the degree of parallelism is significantly increased compared to traditional Householder QR, a large number of flops is still performed in terms of Level-1 and Level-2 BLAS kernels when computing the QR of the local blocks. The distributed version of TSQR is implemented in the SLATE library~\cite{GatesSlate2019} within the Communication Avoiding QR (CAQR) algorithm, where it is used for QR factorisation of tall-and-skinny panels. CAQR was developed for general matrices and not for TS matrices. It is more performant than the standard distributed QR as it replaces the communication volume with more flops.

An alternative to the Householder-based QR algorithms for tall and skinny matrices is the CholeskyQR algorithm \cite{Golub2013MatrixComputations}. CholeskyQR is a simple algorithm with very low communication overhead and about half the arithmetic cost of TSQR. The algorithm is based on Cholesky factorisation and triangular orthogonalisation ($Q = A R^{-1}$) and is generally faster than TSQR. The main drawback is that the algorithm suffers from a loss of orthogonality and is numerically unstable. To compensate for this shortcoming, the authors in \cite{Fukaya2014CholeskyQR2:System, Yamamoto2015RoundoffAlgorithm} have proposed a new algorithm called CholeskyQR2 (see section~\ref{sec:cholqr2} for more details). The main idea is to repeat the CholeskyQR algorithm twice to increase orthogonality. The new algorithm achieves accuracy and orthogonality comparable to TSQR, but also requires twice as much communication as CholeskyQR, with arithmetic cost equivalent to those of TSQR. However, the algorithm fails for input matrices with very high condition numbers ($\kappa(A) > u^{-1/2}$, where $u$ is the unit roundoff). Distributed parallel CholeskyQR is available in the SLATE library, where it is used as one of the preferred methods for solving least squares and QR factorisation problems. However, there is no report on how numerical instabilities have been tackled with ill-conditioned matrices.

The CholeskyQR2 has proven to be the method of choice for solving least squares and eigenvalue problems on highly parallel computer clusters. In~\cite{Hutter2019CACHOLQR2}, the CholeskyQR2 is extended over a 3D processor grid, which results in less communication between processors and outperforms the QR factorisation of ScaLAPACK by up to $3.3\times$ for strong scaling tests and $1.9\times$ for weak scaling tests. However, the main issue with the numerical stability of the algorithm remains. If the condition number of the input matrix is too large, the constructed Gram matrix fails to be positive semidefinite, causing the Cholesky factorisation step to fail~\cite{Fukaya2014CholeskyQR2:System}.

This issue is directly addressed in Shifted CholeskyQR algorithm \cite{Fukaya2020ShiftedCholQR}. The idea is to construct a shifted Gram matrix $\hat{W} := A^T A + \sigma I$ in which the shift factor $\sigma$ ensures the numerical stability of the Cholesky factorisation. However, the proposed solution does not work for matrices with condition numbers close to or greater than $u^{-1/2}$. In the case of extremely ill-conditioned matrices, the Shifted CholeskyQR is used as a preconditioner for CholeskyQR2 and the new algorithm is called Shifted CholeskyQR3. Although this approach significantly improves stability, it leads to $50\%$ more flops compared to CholeskyQR2. 

Another approach to improve the applicability of CholeskyQR for ill-condition matrices is to use the LU decomposition as preconditioning for the Cholesky factorisation~\cite{TeraoLUCholQR2020}. After the LU factorisation $PA := LU$ has been calculated, the Cholesky decomposition can be applied to the matrix $L^T L$, since $L$ is usually better conditioned than $A$. 
After the first call of CholeskyQR with LU, the $Q_1$ obtained is refined by the second call of the standard CholeskyQR algorithm. The achieved orthogonality and the residuals of the algorithm are comparable to those of the Householder QR even for very ill-conditioned matrices, i.e. $\kappa(A) > u^{-1/2}$. However, since the algorithm computes the LU decomposition with partial pivoting, it exhibits less parallelism compared to CholeskyQR2 and is about $1.5$ times slower on shared memory systems and between 3 and 5 times on distributed memory systems.

The most recent improvements to the orthogonality and stability of the CholeskyQR model was proposed in~\cite{Higgins2023-RandHHQR} with the so-called Randomised Householder-Cholesky QR Factorization with Multisketching. The proposed solution introduces up to two randomised sketch matrices (multisketching), which prove that the orthogonality error is bounded by a constant of the order of unit roundoff for matrices of arbitrary condition number. The first step is to compute the randomised Householder QR that generates the matrix $Q_1$ orthogonal to the given sketch matrices. As in other similar approaches, the obtained matrix is reorthogonalised in the second step by calling CholeskyQR2 to produce a fully orthogonal matrix $Q$. 
The authors reported that their approach is applicable to extremely tall-and-skinny matrices ($ n \le 0.01\% m$ ) and was negligibly faster than CholeskyQR2, but more stable than the shifted CholeskyQR3.

Since the orthogonality error of CholeskyQR depends quadratically on the condition number, the authors in~\cite{Yamazaki2015MixedCholeskyQR} have proposed a mixed-precision approach. In this approach, the input and output matrices are retained in their required precision, while certain intermediate results are calculated at doubled precision. The analysis has shown that the orthogonality error of the mixed-precision CholeskyQR approach has a linear dependence on the condition number of the input matrix when doubled precision is used. The main drawback of the proposed algorithm is that the number of floating point instructions increases significantly when doubled precision is used (especially in the case when the target precision is 64-bit).

In \cite{Tomas2019CholeskyProcessors}, the stability of the Cholesky decomposition and orthogonality for very ill-conditioned matrices is restored by combining the CholeskyQR with the Gram-Schmidt method of re-orthogonalisation. A detailed overview can be found in the subsection~\ref{subsec:CholQR2gs}. The proposed algorithm has shown that the obtained orthogonality is equivalent to the Householder-based QR factorisations (general and TSQR), but cannot exploit massively parallel systems as the method uses very tall and skinny column panels with full column size.

The original scientific contributions in this paper are:
\begin{itemize}
 \item A novel algorithm for computing the QR factorisation of ill-conditioned tall-and-skinny matrices which combines the classical CholeskyQR2 algorithm with Gram-Schmidt reorthogonalisation.
 \item Orthogonality and numerical stability of the new algorithm are comparable to those of Householder-based QR factorisation algorithms.
 \item QR factorisation optimised for distributed memory GPU systems.
 \item The algorithm outperforms other methods, including the Shifted CholeskyQR3 algorithm and traditional Householder-based approaches (such as ScaLAPACK's \texttt{PDGEQRF}).
 \item An analysis and synthesis of the state-of-the-art in CholeskyQR-based QR factorisation algorithms for tall-and-skinny matrices.
    
\end{itemize}

When defining the workflow in this paper, our main goal was to guide the readers through all stages of the algorithm development, starting from the basic version and gradually improving it to an advanced version. At each step, we provide a numerical and a performance analysis which helps the reader to better understand the improvements over the previous version. We believe that this approach provides a deeper understanding of our algorithm.
The paper starts with the testing environment, state-of-the-art algorithms and test matrices described in section \ref{sec:stateoftheart}. The CholeskyQR algorithm, a basic building block of our new algorithm, is described in section \ref{sec:cholqr}, followed by the analysis of the CholeskyQR2 algorithm in section \ref{sec:cholqr2}. The main part of the paper is in section \ref{sec:fixCholQR2}, where we describe step by step how the algorithm evolves from the first improvement, shifted CholeskyQR3, through the introduction of the Gram-Schmidt method to our final version, the modified CholeskyQR2 with Gram-Schmidt. The scalability analysis of the new algorithm and the comparison with ScaLAPACK are shown and discussed in section \ref{sec:scalability}.
The paper is concluded in section \ref{sec:conclusion} with a summary of what has been achieved and an overview of future work.
\section{Testing environment}
\label{sec:stateoftheart}

\subsection{Testing platform}
The numerical stability and performance were tested on the Supek supercomputer at the University of Zagreb, University Computing Centre (SRCE). The tests were performed on two partitions: GPU and CPU. The GPU partition consists of 20 nodes connected to the Cray Slingshot interconnect. Each node has an AMD EPYC 7763 CPU with 64 cores and 512 GB of main memory, supported by four NVIDIA A100 Tensor Core GPUs with 40 GB of device memory each. The CPU partition consists of 52 nodes similar to the GPU partition, but with 2 AMD EPYC processors (128 cores in total) and without GPU accelerators. The codes were compiled with the GCC 12.1.0 compiler and the CUDA 11.6 library, which provides BLAS and LAPACK functionality. Cray MPICH 8.1.20 and NVIDIA NCCL 2.12.12 were used for communication. All experiments were performed in double precision arithmetic and an average execution time of 10 runs is reported for each experiment.

The numerical accuracy is obtained by analysing the orthogonality of the obtained matrix $Q$ using the formula $||Q^T Q - I||_F/\sqrt{n}$, where $I$ is the identity matrix, and the residual $||Q R - A||_F/||A||_F$. Both the orthogonality and the residual should be of the order of $O(u)$, where $u$ is a machine precision of a certain numerical type precision (e.g. double precision).

\subsection{Test matrix suite}
\label{subsec:testMatrixSuite}
The input data are artificially generated matrices with a condition number $\kappa (A)$ in the range of $\{10^0,10^1,\ldots,10^{15}\}$. The matrices are generated using the SVD ($U\hat{\Sigma} V$), where U and V are left and right singular vector matrices obtained from the SVD of a random input matrix. The new diagonal matrix $\Sigma$ is constructed so that the diagonal elements are $(1, \sigma^{\frac{1}{n-1}}, \dots, \sigma^{\frac{n-2}{n-1}}, \sigma)$, where the parameter $\sigma$ controls the condition number of the generated matrix $\kappa(A)\approx \sigma$.

For the numerical stability tests, we used matrices of size $30000\times 3000$ and the condition number in the range $\{1, 10^1,\ldots, 10^{15} \}$ in double precision.
The strong tests were performed for matrices with $120000$ rows and the number of columns equal to $1\%$, $5\%$ and $10\%$ of the number of rows. The weak scalability analysis was performed on matrices with $40k, 80k, 120k, \ldots, 480k$ rows and the number of columns fixed to $3000$, resulting in blocks of size $10k \times 3k$ per process (MPI or NCCL rank).

\subsection{Software}
In our tests and analyses, we compare the performance of our novel CholeskyQR2-based algorithms and evaluate their scalability on systems with distributed memory.
To the best of our knowledge, ScaLAPACK is the only publicly available library that can handle QR factorization on distributed memory architectures. The downside is that it does not support execution on GPUs and only implements the Householder-based methods. The only alternative, the SLATE library, which implements CholeskyQR for distributed multi-GPU architectures, exhibits the same numerical instabilities for matrices with very high condition numbers (greater than $10^8$), even when CholeskyQR2 is used. Since it is not possible to compare our solutions with SLATE for ill-conditioned matrices, the algorithm is only compared with ScaLAPACK.

For simplicity and easier comparison with our solution, the total computation and communication costs for ScaLAPACK \texttt{PSGEQRF} implementation are: $2\frac{mn^2}{P}-\frac{2}{3}\frac{n^3}{P}$ (number of operations), $\frac{n^2}{2}\log P$ (number of words transmitted), $2n\log P$ (number of messages), where $m$ and $n$ are the number of rows and columns, respectively, and $P$ is the number of processes.

\section{CholeskyQR}
\label{sec:cholqr}

CholeskyQR is a simple algorithm that calculates the QR factorisation of a tall and skinny matrix. The pseudocode is described in Algorithm \ref{alg:cholqr}. The algorithm starts with the construction of the Gram matrix $A^T A$ (line 1) using matrix-matrix multiplication, after which the upper triangular matrix $R$ is obtained via Cholesky factorization (line 2). Finally, the orthogonal matrix $Q$ is constructed by right-multiplying $A$ with the upper triangular matrix $R$ (line 3), e.g. using the routine triangular system solve (\texttt{trsm}) from the LAPACK library.
The main advantage of CholeskyQR over Householder-based algorithms such as TSQR is that all steps in Algorithm~\ref{alg:cholqr} can be implemented as Level-3 BLAS operations, which ensures significantly better performance on large parallel systems.

\begin{algorithm}[h]
  \caption{CholeskyQR}\label{alg:cholqr}
  \begin{algorithmic}[1]
    \Require {$A \in \mathbb{R}^{m\times n}$}
    \Ensure {$Q \in \mathbb{R}^{m\times n} $ orthogonal and $ R \in \mathbb{R}^{n\times n}$ upper triangular matrix}

    \State $W := A^T A$ \Comment{Construct Gram matrix}
    \State $W = R^T R$ \Comment{Cholesky factorization}
    \State $Q := A R^{-1}$  \label{alg:cholqr:q}
    
  \end{algorithmic}
\end{algorithm}

\textbf{\textit{Parallelization.}} 
Since the CholeskyQR is intended for factorization of matrices with many more rows than columns, parallelization is done via the row dimension.
The matrix $A$ is divided into one-dimensional row blocks (see Fig.~\ref{fig:distA} middle ), with each processor $i$ processing a row block $A_i$. The Gram matrix is first constructed locally, with each processor calculating its local matrix $W_i = A^T_i A_i$ ( lines 1--3). Then, a collective communication is required to collect and sum all local components to obtain the final Gram matrix $W = W_1 + W_2 + \ldots + W_P$ (line 4). The Cholesky factorization (line 5) is performed redundantly by each processor using a highly optimized implementation designed for shared memory systems, e.g. from the MAGMA or LAPACK library. Finally, the construction of the orthogonal matrix $Q$ (lines 6--7), can be done in parallel as each processor constructs its own part of $Q$, so that no further communication is required. Overall, CholeskyQR requires only a single collective communication.

If the matrix $Q$ is required in a single memory space (e.g. a compute node), an additional global communication (e.g. ~\texttt{MPI\_gather}) is required to collect the blocks $Q_i$ distributed across the MPI ranks. In the remainder of the paper, we assume that $A$ is already distributed across the processors and $Q$ does not need to be collected on a single MPI rank. This simulates the real-world scenario where QR factorization is only a part of a larger computer code where $A$ is already distributed and $Q$ needs to be distributed for further processing (see~\cite{Winkelmann2019}).
Throughout the paper, the parallel version of CholeskyQR (CQR) is used as the main building block for all other algorithms.

\begin{algorithm}[h]
  \caption{Parallel CholeskyQR (CQR)}\label{alg:cholqr_distributed}
  \begin{algorithmic}[1]
    \Require {$A \in \mathbb{R}^{m\times n}$, $P$ number of processors}
    \Ensure {$Q \in \mathbb{R}^{m\times n} $ orthogonal and $ R \in \mathbb{R}^{n\times n}$ upper triangular matrix}

    \For {$ i \gets 1,P$}           \Comment{Parallel \texttt{for}}
        \State $W_i := A_i^T A_i$    \Comment{Compute Gram locally}
    \EndFor
    \State $W := \sum_i W_i$ \Comment{\texttt{Allreduce}}
    \State $W = R^T R$ \Comment{Cholesky factorization}
    \For {$ i \gets 1,P$}           \Comment{Parallel \texttt{for}}
        \State $Q_i := A_i R^{-1}$  \Comment{Compute Q locally}
    \EndFor
  \end{algorithmic}
\end{algorithm}

The computational cost (flops) of parallel CholeskyQR is $\frac{1}{3} n^3 + 2 \frac{m}{P} n^2 + n^2 \log_2 P$ with the dominant factor  $\left(2\frac{m}{P}n^2\right)$, coming from the construction of the Gram matrix (gemm/syrk) and the construction of $Q$ (trsm). The total cost of CQR is about half of the flops required by TSQR. As described in Algorithm~\ref{alg:cholqr_distributed}, CholeskyQR can be implemented with only one call to a collective communication routine, for example \texttt{Allreduce} (Algorithm~\ref{alg:cholqr_distributed}, line 4), which makes CQR suitable for execution on large parallel systems. The number of words transmitted and the number of messages is $n^2 \log_2 P$ and $\log_2 P$ respectively and corresponds to the transmission (broadcasting and reduction) of the local matrices $W_i$ and is the same (except for a constant factor) to that of TSQR. More details on the communication and calculation complexities can be found in Table~\ref{tab:cholqr2} (under CQR).

The biggest disadvantage is that the CholeskyQR is numerically unstable. The loss of orthogonality of the calculated $Q$ increases with the condition of the matrix $A$ and is upper bounded by $\mathcal{O}(\epsilon\kappa(A)^2)$~\cite{StathopoulosBlockOrtho2006}, where $\epsilon$ is the machine precision ($\approx 10^{-16}$). Even for small condition numbers, e.g. $\kappa(A) = \mathcal{O}(10)$, the deviation from orthogonality obtained with double precision arithmetic is $\mathcal{O}(10^{-14})$. Furthermore, since the construction of the gram matrix squares the condition number, the resulting $W$ is not positive semidefinite for ill-conditioned matrices, so the Cholesky factorisation step cannot be performed. 

\section{CholeskyQR2}
\label{sec:cholqr2}

As already shown, the CholeskyQR cannot always generate orthogonal vectors ($Q$), and as the condition number increases, it even becomes numerically unstable, which often leads to the failure of the algorithm (reduced stability of the Cholesky factorisation). To address the problem with orthogonality, a simple and very effective idea was presented to reorthogonalise the obtained $Q$ by repeating the CholeskyQR algorithm (see \cite{Fukaya2014CholeskyQR2:System}). The proposed algorithm is called CholeskyQR2.

The pseudocode for the CholeskyQR2 is given in Algorithm \ref{alg:cholqr2}.
The algorithm starts with the calculation of the QR factorization of $A$ (line 1), for which the Cholesky QR algorithm (CQR) is used. The result is the matrix $Q_1$, which does not have to be orthogonal. To improve the numerical accuracy of the obtained matrix, another orthogonalization step is performed on $Q_1$ ( line 2). The idea of repeating the orthogonalization was introduced in \cite{Giraud2005RoundingProcess}, where the authors showed that the stability of the Gram-Schmidt algorithm and the orthogonality of the computed vectors can be improved if the algorithm is executed twice. The idea is also applicable to the Cholesky QR, since both algorithms are of the triangular orthogonalization type, i.e. the factor $Q$ is calculated by right multiplication of the triangular matrix $R$ ( Algorithm \ref{alg:cholqr} line 3). Finally, the triangular matrix $R$ is obtained by multiplying the upper triangular factors (line 3), which are generated by two calls to CholeskyQR.

\begin{algorithm}[h]
  \caption{CholeskyQR2 (CQR2)}\label{alg:cholqr2}
  \begin{algorithmic}[1]
    
    \Require{$A \in \mathbb{R}^{m\times n}$}
    \Ensure{$Q \in \mathbb{R}^{m\times n}$ orthogonal and $R \in \mathbb{R}^{n\times n}$ upper triangular matrix}

    \State $ [Q_1, R_1] := CQR(A)$
    \State $ [Q, R_2] := CQR(Q_1) $
    \State $R := R_2 R_1$
    
  \end{algorithmic}
\end{algorithm}

Although repeating CholeskyQR twice significantly improves orthogonality, CholeskyQR2 is still numerically unstable for ill-conditioned matrices with condition numbers greater than $\mathcal{O}(u^{-1/2})$, where $\mathbf{u}$ is the unit roundoff.
In practice, this means that the algorithm is stable up to condition numbers of $10^8$~\cite{Fukaya2014CholeskyQR2:System, Yamamoto2015RoundoffAlgorithm} and the deviation from orthogonality of the computed $Q$ is of order $\mathcal{O}(\kappa(A)^2 u)$.

\begingroup
\renewcommand{\arraystretch}{1.5} 
\begin{table}[]
\caption{Computational (comp) and communication (comm) costs of routines in parallel CholeskyQR
and the CholeskyQR2 algorithms. We assume that the routine \texttt{syrk} is used to compute the Gram matrix, and not \texttt{gemm}.}
\resizebox{0.5\textwidth}{!}{%
\label{tab:cholqr2}
\begin{tabular}{ccccc}
\toprule
\multicolumn{2}{c}{algorithm} & routine & comp & comm \\
\cmidrule(lr){1-2} \cmidrule(lr){3-3} \cmidrule(lr){4-4} \cmidrule(lr){5-5} 

\multirow{6}{*}{CQR2} & \multirow{5}{*}{CQR} & Gram & $\frac{m}{P} n^2$ & -\\
 &  & Gram\_reduce & $n^2\log_2P$ & $n^2\log_2P$ \\
 &  & Cholesky & $\frac{n^3}{3}$ & - \\
 &  & Construct\_Q & $\frac{m}{P}\ n^2$ & - \\  \cmidrule(lr){3-5}
 &  & Total & $\frac{1}{3}n^3 + 2\frac{m}{P}\ n^2 + n^2 \log_2P$ & $n^2 \log_2P$ \\
 \cmidrule(lr){3-5}
 & Compute\_R &  & $\frac{1}{3}\ n^3$ & - \\
 
\cmidrule(lr){2-5} 
& Total &  & $n^3 + 4\frac{m\ n^2}{P} + 2n^2\log_2P$ & $2\ n^2\log_2P$  \\
\bottomrule
\end{tabular}
}
\end{table}
\endgroup

The parallelization of the algorithm is simple as it exploits the parallelism within the CholeskyQR algorithm. Therefore, no additional communication is required, except for two collective routines, one for each call to CholeskyQR. Since the algorithm repeats CholeskyQR twice, the communication costs (see Table~\ref{tab:cholqr2} CQR2) are twice as high as for CholeskyQR and amount to $2n^2 \log_2 P$. Although the communication costs are much higher than for TSQR, they are still much lower than for Householder QR. The computational cost is also twice that of CholeskyQR, plus an additional cost of $\frac{1}{3}n^3$ incurred by the construction of the matrix $R$ (Algorithm~\ref{alg:cholqr2} line 3). The total computational cost of CholeskyQR2 is comparable to that of TSQR.

\section{CholeskyQR variants for extremely ill-conditioned matrices}
\label{sec:fixCholQR2}

The source of the numerical instability in the CholeskyQR algorithm is the Cholesky decomposition, which is sensitive to nearly singular or ill-conditioned matrices. Such matrices can occur in the construction of the Gram matrix whose condition number is a squared condition of the original matrix, resulting in a matrix that is not semi-positive definite in finite precision arithmetic. In the following subsections, we present recent developments that address the critical challenge of the breakdown of Cholesky factorization within CholeskyQR algorithms and introduce our new parallel CholeskyQR-based algorithm for distributed memory systems. These novel algorithms provide innovative strategies to circumvent the limitations imposed by ill-conditioned matrices and ensure the stability and reliability of the Cholesky factorization process even for extremely ill-conditioned matrices.

\subsection{Shifted CholeskyQR3}
\label{subsec:sCholQR3}

\begin{figure}[ht]
\centering
\includegraphics[width=0.5\textwidth]{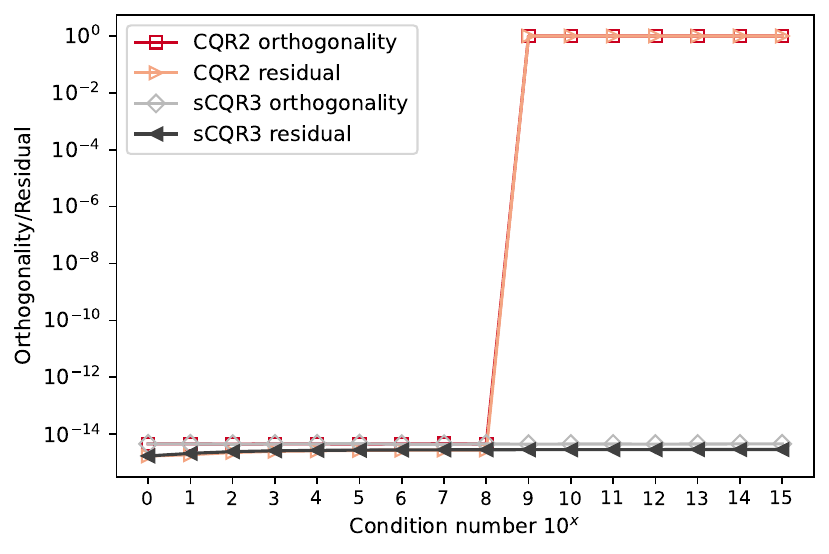}
\caption{Orthogonality and residuals of sCQR3 and CQR2 as a function of the condition number, for input matrices with $m=30000$, $n=3000$ and conservative shift for sCQR3.}
\label{fig:orthoscqr3}
\end{figure}

A recent proposal to improve the numerical stability of CholeskyQR is to shift the Gram matrix in the CholeskyQR routine~\cite{Fukaya2020ShiftedCholQR}. The main idea is to decrease the condition number of the computed Gram matrix thus improving the stability of the Cholesky factorization. The algorithm is called {\it Shifted Cholesky QR}, and we will refer to it as \textit{sCQR} throughout the paper. The algorithm presented in~\cite{Fukaya2020ShiftedCholQR} is described in Algorithm~\ref{alg:scholqr} and is similar to CholeskyQR except for the key steps of introducing the shift $s$ in lines 2--3. The shift is chosen to force the Gram matrix to become positive definite so that the Cholesky algorithm can be completed. With a well-chosen shift, sCQR is suitable for matrices of condition number up to $\mathcal{O}(u^{-1})$. In our research, we choose the conservative approach (as proposed in~\cite{FukayaPerformanceCholQR2018}), where the Frobenius norm is used instead of norm-2, because of significantly less computational cost, which ensures numerical stability for our test matrices (see Fig.~\ref{fig:orthoscqr3}).

\begin{algorithm}[ht]
  \caption{Shifted Cholesky QR (sCQR)}\label{alg:scholqr}
    \begin{algorithmic}[1]
      \Require{$A \in \mathbb{R}^{m\times n}$}
      \Ensure{$ Q \in \mathbb{R}^{m\times n} $ orthogonal and $ R \in \mathbb{R}^{n\times n} $ upper triangular matrix }
        \State $G = A^T\ A $  
        \State $s = \sqrt{m} \, \mathbf{u} \, ||A||^2_F$ \Comment{calculate shift}
        \State $W = G + sI$ \Comment{construct Gram matrix}
        \State $W = R^T\ R$ \Comment{Cholesky factorization}
      \State $Q=AR^{-1}$
  \end{algorithmic}
\end{algorithm}

The numerical stability for extremely ill-conditioned matrices (e.g. $\kappa(A) \ge 10^{15}$) can be improved by choosing a better shift (usually a smaller shift of the order of $\mathcal{O}(u\| A\|_2$), which can ensure the stability of the Cholesky factorization. However, experimenting with other approaches to compute the shift is not the focus of this article as it does not affect the overall computational complexity or execution time. How to choose an optimal shift for a given matrix can be read in~\cite{Fukaya2020ShiftedCholQR}.

The condition number of the computed matrix $Q$ in Algorithm~\ref{alg:scholqr} (line 5) in the shifted CholeskyQR is roughly upper bounded by $\mathcal{O}(u^{-1/2})$, which is a condition number that ensures the numerical stability of the CholeskyQR2.
To obtain the orthogonality and the residual of order $u$, the authors in~\cite{Fukaya2020ShiftedCholQR} have proposed a solution where the shifted Cholesky QR is used as a preconditioner for the CholeskyQR2 that ensures its numerical stability. The combined application of shifted CholeskyQR and CholeskyQR2 constitutes {\it shifted CholeskyQR3} (sCQR3), which is described in Algorithm \ref{alg:scholqr3}. The first step (line 1) is to compute the matrix $Q_1$ with the reduced condition number, which is then further orthogonalized via the CholeskyQR2 (line 2). Finally, the resulting matrix $R$ (line 4) by multiplying intermediate triangular factors computed by the shifted CholeskyQR and CholeskyQR2.

\begin{algorithm}[ht]
  \caption{Shifted Cholesky QR3 (sCQR3)}\label{alg:scholqr3}
    \begin{algorithmic}[1]
      \Require{$A \in \mathbb{R}^{m\times n}$}
      \Ensure{$ Q \in \mathbb{R}^{m\times n} $ orthogonal and $ R \in \mathbb{R}^{n\times n} $ upper triangular matrix }
      \State $[Q_1, R_1] = sCQR(A)$
      \State $[Q, R_2] = CQR2(Q_1)$
      \State $R:= R_2R_1$
  \end{algorithmic}
\end{algorithm}

The parallelization of the shifted CholeskyQR routine is similar to that of the CholeskyQR routine described in Section \ref{sec:cholqr}. The algorithms are the same except for the computation of the shift factor and shifting of the Gram matrix. The main additional cost is the calculation of the Frobenius norm of the matrix in computing the shift (Algorithm~\ref{alg:scholqr} line 2). This is done by summing over the squares of the elements of the partial matrix stored in each rank and communicating the results to rank 0, where the norm is calculated. Based on the obtained results, the shift is added to the partial Gram matrix stored in rank 0. The total computational cost of the parallel shifted CholeskyQR3 with P processors is:
\begin{equation}
    \frac{5}{3} n^3 + 6 \frac{m\ n^2}{P} + 3\ n^2\ log_2P + 2\frac{m n}{P}
\end{equation}
where the last term accounts for the calculation of the Frobenius norm when computing the shift. In the above expression, we have neglected inexpensive operations such as the calculation and addition of the shift parameter from the norm, which cost $\mathcal{O}(n)$ or less.
The computational cost of the shifted CholeskyQR3 is larger than that of CQR2, for an additional cost of sCQR. Since the cost of sCQR is about the same as that of CQR, the sCQR3 is about $1.5\times$ higher than CQR2 plus the additional cost ($n^3$) required to compute the final $R$. The communication cost is $50\%$ higher than for cQR2.

With the right choice of shift, the numerical stability of the Cholesky factorization can be ensured, even for the extremely ill-condition matrices. However, to compute the orthogonal matrix $Q$, comparable to the Householder-based approaches, e.g.~TSQR, additional computational steps are required compared to CholeskyQR2. These additional steps significantly increase both computational cost and total execution time, making the shifted CholeskyQR3 inferior in terms of performance.

\subsection{CholeskyQR2 with Gram-Schmidt}
\label{subsec:CholQR2gs}

Another approach to address the problem with instability of the Cholesky factorization for matrices with a condition number greater than $\mathcal{O}(u^{-1/2})$ ($10^8$) is by combining the CholeskyQR2 with the modified Gram-Schmidt re-orthogonalisation \cite{Tomas2019CholeskyProcessors}. The basic version of the algorithm proposed in~\cite{Tomas2019CholeskyProcessors} (without additional look-ahead optimizations) is shown in Algorithm \ref{alg:cholqrgs}. 

To make understanding more easy, the following notation is used in the rest of the paper. In the description of the algorithms, the $A_j$ denotes the $j$-th panel of the matrix $A$ with full row rank, and $A_{i:j}$ stands for the range of panels from the $i$-th to the $j$-th with full row rank. In addition, the term $A_{i,j}$ denotes the tile of dimension $(b\times b$), where $b$ is the width of the panel. 

\begin{algorithm}[ht]
  \caption{Cholesky QR with Gram-Schmidt (CQRGS)}\label{alg:cholqrgs}
    \begin{algorithmic}[1]
      \Require{$A \in \mathbb{R}^{m\times n}$, panel width $b$ and number of panels $k=\frac{n}{b}$}
      \Ensure{$ Q \in \mathbb{R}^{m\times n} $ orthogonal and $ R \in \mathbb{R}^{n\times n} $ upper triangular matrix }
      \For {$j = 1\ldots k$}
        \State $W_j := A_j^T A_j$ \Comment{Construct Gram matrix}
        \State $W_j = U^T U$ \Comment{Cholesky factorization}
        \State $Q_j = A_j U^{-1}$
        \State $R_{j,j} = U$
        \State $Y := Q^T_j A_{j+1:k}$   
        \State $A_{j+1:k} := A_{j+1:k} - Q_j Y$ \Comment{Update panels }
        \State $R_{j,j+1:k} := Y$
    \EndFor
  \end{algorithmic}
\end{algorithm}

The matrix $A$ is processed by panels, where $k$ is the number of panels and $b$ is the panel width (line 1). The first step of each iteration is to compute the Gram matrix (line 2), and the Cholesky factorization (line 3) of the current panel $A_j$, followed by the construction of the orthogonal matrix $Q_j$ (line 4), as in the CQR. As defined by the Gram-Schmidt process, the rest of the panels to the right of the current panel are re-orthogonalized (lines 6--7) w.r.t the $Q_j$ by applying the orthogonal projection $Q_j Q^T_j$ to $A_{j+1:k}$. 
The advantage of the algorithm is that all steps, except the Cholesky factorization in line 3, can be realized using efficient level-3 BLAS kernels.
In its original version~\cite{Tomas2019CholeskyProcessors}, the algorithm was developed and tested for shared memory and hybrid CPU-GPU systems, where the Cholesky decomposition is performed on the CPU while the trailing updates of $A$ are performed on the GPU. Since the algorithm operates on panels with full row rank, parallelism was exploited only in the column (panel) direction, depending on the high-performance Level-3 BLAS kernels used for updating the trailing sub-matrices.

In the rest of this section, we describe our distributed CholeskyQR2 algorithm with the modified Gram-Schmidt re-orthogonalization based on the algorithm presented in~\cite{Tomas2019CholeskyProcessors}. Our algorithm is similar to the original CholeskyQR2 algorithm (see Algorithm \ref{alg:cholqr2}), except that instead of calling CholeskyQR (CQR) twice, our approach involves calling the CholeskyQR with Gram-Schmidt (CQRGS) twice (see Algorithm \ref{alg:cholqr2gs}).

\begin{algorithm}[h]
  \caption{CholeskyQR2 with Gram-Schmidt (CQR2GS)}\label{alg:cholqr2gs}
  \begin{algorithmic}[1]
    
    \Require{$A \in \mathbb{R}^{m\times n}$}
    \Ensure{$Q \in \mathbb{R}^{m\times n}$ orthogonal and $R \in \mathbb{R}^{n\times n}$ upper triangular matrix}

    \State $ [Q_1, R_1] := CQRGS(A)$
    \State $ [Q, R_2] := CQRGS(Q_1) $
    \State $R := R_2 R_1$
    
  \end{algorithmic}
\end{algorithm}

The main contributions are extending the CholeskyQR and Gram-Schmidt (as presented in~\cite{Tomas2019CholeskyProcessors}) to distributed memory systems equipped with GPU accelerators and making some additional optimization that will make the algorithm not only numerically stable for extremely ill-condition problems but also in some cases superior in terms of performance compared to the original CholeskyQR2.

Similar to CholeskyQR, the input matrix $A$ is partitioned and distributed among $P$ processors in a one-dimensional block row layout (see Fig.~\ref{fig:distA} middle). Each block row $A_{p}$ is locally partitioned into $k$ panels $A_{p,j}$ with width $b= n/k$ columns and $j \in \{1,\ldots,k\}$. The proposed block row partitioning allows for coarse-grained parallelization between processors (e.g. compute nodes) while partitioning into panels exploits fine-grained parallelization at the node level.

\begin{figure}[ht]
\centering
\includegraphics[width=0.4\textwidth]{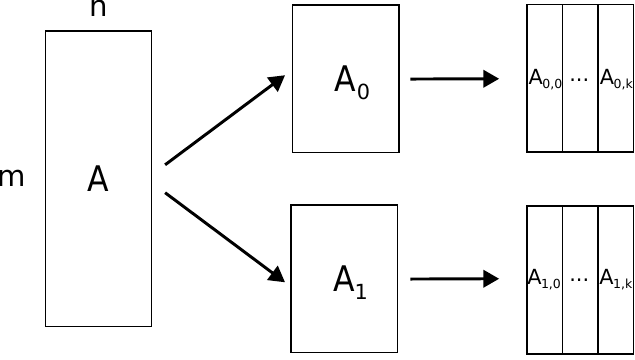}
\caption{Distributing and slicing of a matrix. An example with 2 processors. The assignments of blocks and panels with processors are indicated on the vertical axis.} \label{fig:distA}
\label{fig:distA}
\end{figure}

Our distributed CholeskyQR algorithm with the blocked Gram-Schmidt re-orthogonalization is described in Algorithm \ref{alg:cholqrgs_distributed}.
The algorithm processes the input matrix $A$ by panels, starting from the leftmost one. The first step is to calculate the Gram matrix. To do this, each processor computes a local Gram matrix from its current panel $A_{p,j}$ (line 2) and sums over all the local matrices to get the final Gram matrix (line 3). Summing and distributing requires collective communication (e.g.~\texttt{MPI\_Allreduce}), after which all processors have identical Gram matrix. The Cholesky factorization (line 4) is computed redundantly on each processor, and to avoid unnecessary communication, each processor updates only its block row $Q_{p,j}$ of the orthogonal panel $Q_j$ (line 5). Updating the panels right to the current panel requires communication between the processors since the panel $Q_j$ is applied to the full row rank of the trailing matrix $A_{p,j+1:k}$ (lines 7--9). Each processor first calculates its partial product $Y_p$ (line 7), and then a collective communication (line 8) is required to sum and broadcast the global $Y$. Once the intermediate matrix $Y$ has been transmitted, each block row can be updated independently (line 9). 

\begin{algorithm}[ht]
  \caption{Distributed Cholesky QR with blocked Gram-Schmidt}\label{alg:cholqrgs_distributed}
  \begin{algorithmic}[1]
  \Require{Number of processors $P$, $A \in \mathbb{R}^{m\times n}$ partitioned into block rows and distributed among processors, panel width $b$}
  \Ensure{$ Q \in \mathbb{R}^{m\times n} $ orthogonal and $ R \in \mathbb{R}^{n\times n} $ upper triangular matrix}
  \For{j=1,2, \dots, k}
    \State $W_{p, j} := A_{p,j}^T A_{p,j}$
    \State $W_j := \text{MPI\_Allreduce}(W_{p, j})$ \Comment{Communication}
    \State $W_j = U^T U$
    \State $Q_{p,j} := A_{p,j} U^{-1}$
    \State $R_{j,j} := U$
    \State $Y_p := Q_{p,j}^T [A_{p, j+1}, A_{p, j+2}, \dots , A_{p, k}]$
    \State $Y := \text{MPI\_Allreduce}(Y_p)$ \Comment{Communication}
    \State $[A_{p,j+1}, \dots, A_{p,k}] := [A_{p,j+1}, \dots, A_{p, k}] - Q_{p,j} Y$
    \State $[R_{j,j+1}, R_{j,j+2}, \dots, R_{j,k}] := Y$
    \EndFor
  \end{algorithmic}
\end{algorithm}

A special case in which $b := n$, leads to a single large panel eliminating the iteration over $j$ and the panel updates, effectively removing the Gram-Schmidt from the algorithm. In this case, the CQR2GS falls back to CholeskyQR2, resulting in computational and communication costs equivalent to those of CholeskyQR2.

A detailed breakdown of the computational and communication complexity for each subroutine can be found in Table \ref{tab:cqr2gs}. As shown in the table, the total complexity cost of the parallel CholeskyQR2 with Gram-Schmidt (CQR2GS) is dominated by expression $4\frac{m\ n^2}{P}$ that arises from the total cost of routines \texttt{syrk}, \texttt{trsm} and \texttt{update}, that correspond to the construction of Gram matrix, computing orthogonal matrix and updating part of Gram-Schmidt process, respectively. Since it does not depend on the panel width $b$, the computational complexity for these parts can easily be decreased by using more processors and exploiting more parallelism.
Opposite of that, the first and last operands in the CQR2GS total computational costs, which correspond to the Cholesky factorization and the matrix addition performed in the reduction calls, depend on the panel width $b$. Therefore, $b$ can be considered as an optimization parameter that can be adjusted to achieve either improved numerical results, such as orthogonalization of the matrix, or optimized computational performance by reducing the computational complexity. This approach, characterized by a parameter $b$, not only increases the numerical stability but also reduces the overall complexity compared to the conventional CholeskyQR2 algorithm.

The reason why the numerical stability depends on the value of $b$ lies in the fact that when input matrix $A$ is divided into panels, the condition number of the first panel can be significantly decreased. Since the condition number is the ratio between the largest and the smallest singular value, we have to assess the upper and lower bounds of the singular values of the submatrix (i.e.~panel) $B \in \mathbb{R}^{m\times r}$ of the starting matrix $A \in \mathbb{R}^{m\times n}$ with $n-r$ columns removed and $r < n$.

Let $\sigma_1 \ge \sigma_2 \ge \ldots \ge \sigma_{n-1} \ge \sigma_n$ be the singular values of $A$ and $\gamma_1 \ge \gamma_2 \ge \ldots \ge \gamma_{r-1} \ge \gamma_r$ singular values of $B$, then the following inequalities hold:
\begin{align}
    \sigma_i &\ge \gamma_i, &\;   &i = 1,2,\ldots,r \\
    \gamma_i &\ge \sigma_{i+(n-r)}, &\;  &i \le r.
\end{align}

\noindent
For a proof see Colloraly 3.8.6 in~\cite{Golub2013MatrixComputations}. From the above inequality, the largest and the smallest singular values of $B$ are bounded by:
\begin{align}
    \sigma_1 &\ge \gamma_1 \ge \sigma_{1+(n-r)} \\
    \sigma_r &\ge \gamma_r \ge \sigma_{n},
\end{align}
and the condition number of $B$ is then:
\begin{equation}\label{eq:cond_bounds}
cond(A) = \frac{\sigma_1}{\sigma_n} \ge cond(B) \ge \frac{\sigma_{1+(n-r)}}{\sigma_r}.
\end{equation}
In the worst case, the left equality holds and the condition number of the panel is equal to the condition of the original matrix. This is the case of very clustered singular values with an extremely large singular value where the CQRGS (CQR2GS) cannot secure numerical stability. However, such extreme use-cases are not the focus of this research and are the topic for future work. In the general use-case, the condition number of $B$ is much smaller than the upper bound, especially when the singular values of the original matrix $A$ are equidistantly distributed.
On this basis, the split of the matrix, i.e. the panel width $b$ can be chosen so that the condition number of the first panel $A_{1}$ is small enough so that its Gram matrix $A_{1}^T A_{1}$ has a condition number less than $\mathcal{O}(u^{-1})$ ($10^{15}$ in the case of double precision arithmetic).

\begingroup
\renewcommand{\arraystretch}{1.5}
\begin{table*}[]
\caption{Computational (up) and communication (down) complexity of CholeskyQR2 with Gram-Schmidt (GS).}
\label{tab:cqr2gs}
\centering
\begin{tabular}{ccccc}
\toprule
\multicolumn{2}{c}{\textbf{algorithm}} & \textbf{routine} & \textbf{total comp} & \\
\cmidrule(lr){1-5}
\multirow{9}{*}{CQR2GS} & \multirow{6}{*}{CQRGS} & Gram & $b\ n\ \frac{m}{P}$ &  \\
 &  & Gram\_reduce & $b\ n\ log_2P$ &  \\
 &  & Cholesky & $\frac{b^2\ n}{3}$ &  \\
 &  & Construct\_Q & $\frac{b\ m\ n}{P}$ &  \\
 &  & GS & $2\frac{m\ n}{P}(n-b)$ &  \\
 &  & GS\_reduce & $\frac{n}{2}(n-b)\ log_2P$ & \\
 \cmidrule(lr){3-5}
 &  & Total & $\frac{b^2\ n}{3} + 2\frac{m\ n^2}{P} + \frac{n}{2}(n+b)\ log_2P$ &  \\
 \cmidrule(lr){3-5}
 & Compute\_R &  & $\frac{n^3}{3}$ &  \\ 
 \cmidrule(lr){2-5}
 & Total &  & $2\frac{b^2\ n}{3} + \frac{n^3}{3} + 4\frac{m\ n^2}{P} + n(n+b)\ log_2P$ &  \\
\bottomrule
 &  & & \textbf{total comm} & \textbf{\# of calls} \\
\cmidrule(lr){4-5}
\multirow{3}{*}{CQR2GS} & \multirow{2}{*}{CQRGS} & Gram\_reduce & $b\ n\ log_2P$ & $\frac{n\ (n+b)}{2b^2}$ \\
 &  & GS\_reduce & $\frac{1}{2} n(n-b)\ log_2P$ & $\frac{n\ (n-b)}{2b^2}$ \\
 \cmidrule(lr){2-5}
 & Total &  & $n\ (n+b)\ log_2P$ & $2\frac{n^2}{b^2}$ \\
\bottomrule
\end{tabular}
%}
\end{table*}
\endgroup

\begin{figure}[ht]
\centering
\includegraphics[width=0.45\textwidth]{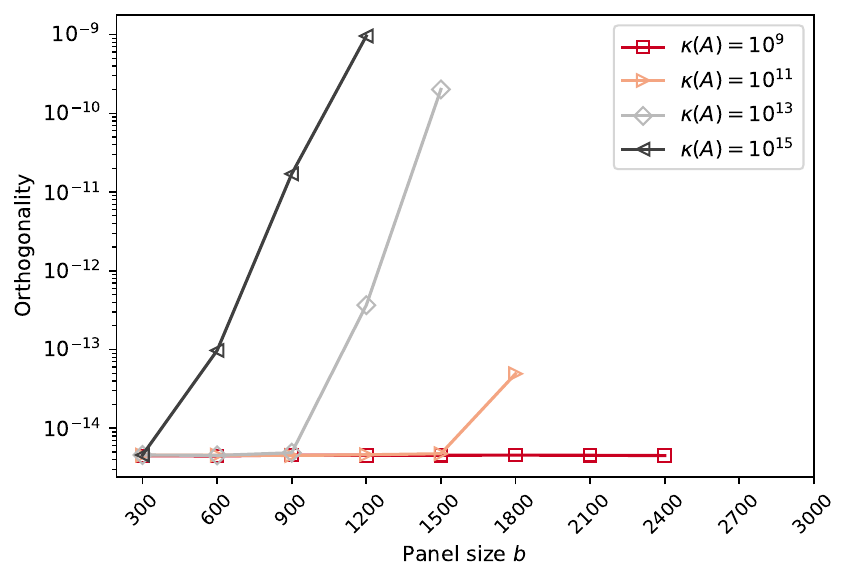}
\caption{CQR2GS: Orthogonality of $Q$ as a function of panel size, for ill-condition input matrices with $m=30000$, $n=3000$.}
\label{fig:cqr2gs_ortho_b}
\end{figure}

Too large $b$ can lead to a large condition number of the Gram matrix $W_j = A_j^T A_j$ and to numerical instability of the Cholesky factorization (see Fig. \ref{fig:cqr2gs_ortho_b}). For an ill-conditioned matrix with $\kappa(A) = 10^\textbf{15}$, the block size should be very small ($b = 300$) to achieve the orthogonality of order of ${u}$ resulting in dividing the matrix into a larger number of panels (concretely 10) and iterating over lines 2--10 (Algorithm~\ref{alg:cholqrgs_distributed}, line 1). On the other hand, for smaller condition numbers, the panel width is increased and the number of iterations is decreased. Note that in the case of matrices with condition numbers smaller than $O(10^8)$, only one panel is needed and CholeskyQR with Gram-Schmidt becomes a standard CholeskyQR.

\begin{figure}[ht]
\centering
\includegraphics[width=0.5\textwidth]{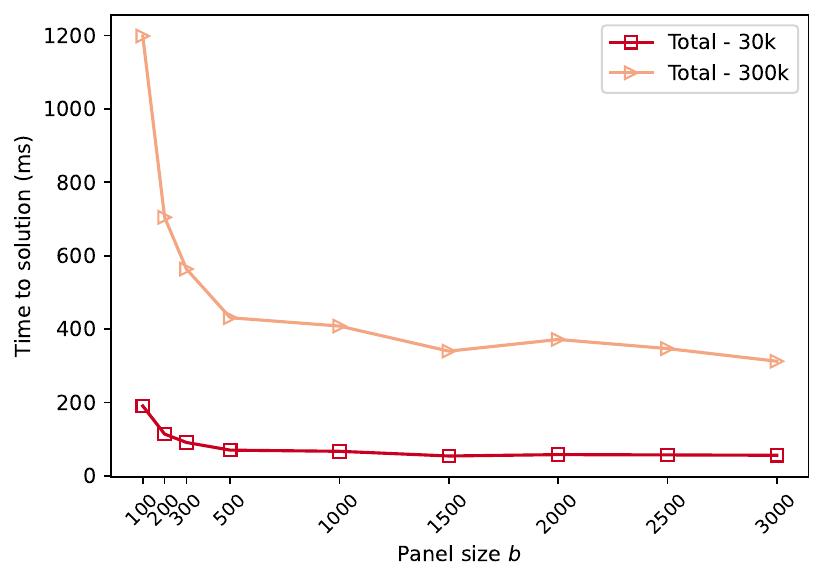}
\caption{Time to solution of CQR2GS on 4 GPUs as a function of panel size for well-condition input matrices ($\kappa(A) = 10^4$) with $m=\{30000, 300000\}$ and a fixed number of columns $n=3000$.}
\label{fig:timeexecqr2gs}
\end{figure}

The connection between the number of panels and the performance (total execution time) is shown in Fig.~\ref{fig:timeexecqr2gs}. It is observed that the panel width $b$ has a significant impact on the performance of the parallel CQR2GS and that for a larger panel width (i.e.~a smaller number of panels) the execution time decreases. By decreasing the panel width (i.e.~increasing the number of panels), both the number of flops and the communication (volume of data) decreases (see total \textit{flops}, \textit{comm} and \textit{\#calls} in Table~\ref{tab:cqr2gs}). However, with a larger number of panels, the number of communication calls significantly increases (with the factor $n^3 / 2)$, which results in a significantly longer total execution time (the left side in Fig~\ref{fig:timeexecqr2gs}.

Another advantage of the CholeskyQR2 with Gram-Schmidt is the reduced amount of communicated data, resulting in lower communication intensity compared to CholeskyQR2. The total communication cost of the distributed CQRGS is:
\begin{equation} 
\label{eq:cholqr2_cost}
    n\ (n+b) \Log_{2}P,
\end{equation}
compared to the $2 n^2 \Log_{2}P$ communication cost of CholeskyQR2. Although additional communication is introduced by Gram-Schmidt process (see Algorithm~\ref{alg:cholqrgs_distributed}, line 8), a much significant communication decrease is achieved in constructing Gram matrix with smaller $b$ resulting in a smaller Gram matrix that has to be reduced and broadcasted among all MPI ranks. This expression highlights that the volume of communicated data remains consistently lower than that of the CholeskyQR2 algorithm as far as $b << n$. 

It is worth noting that although a smaller $b$ contributes to a reduction in total costs, it leads to a higher number of memory operation calls 
$\left(2\frac{n^2}{b^2}\right)$ 
if a larger number of panels is used, which results in higher overhead. If the panel width $b$ is halved, the number of communication messages increases by $4$ times. 
Despite the reduced flops count for a smaller $b$, the influence of the communication on the total execution time is more dominant than the improvements achieved by decreasing the number of flops as illustrated in Fig.~\ref{fig:timeexecqr2gs}. Although the flops count decreases with the smaller panel width (e.g. $b=100$), the total execution time is much longer than if two panels are used (e.g.~$b=1500).$ 

This highlights the nuanced optimization required in the selection of $b$ to balance computational efficiency, communication overhead, and numerical results. 
According to Fig.~\ref{fig:cqr2gs_ortho_b}, for our specific input matrices, the algorithm requires 10 panels to maintain orthogonality at a satisfactory level for input matrices with condition number $10^{15}$. As a result, the algorithm itself does not work in optimal mode. In this context, it would be beneficial to investigate a modification that still works with good numerical behavior, but with slightly larger values of $b$.

\subsection{Modified CholeskyQR2 with Gram-Schmidt}
\label{subsec:mCholQR2gs}

\begin{figure*}[t]
\centering
\includegraphics[width=0.8\textwidth]{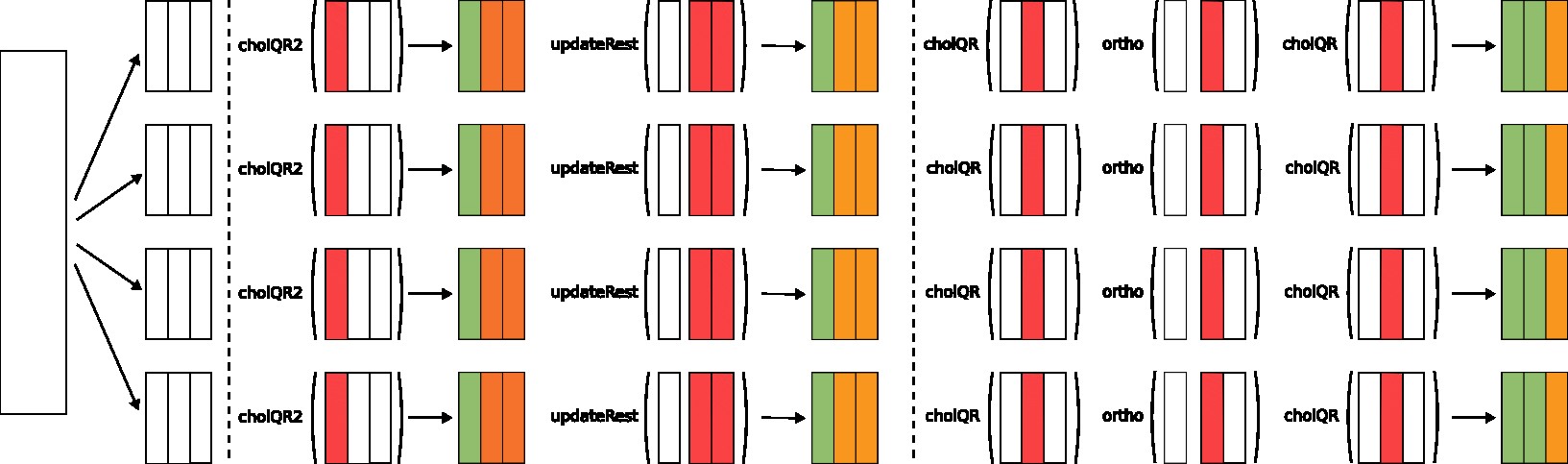}
\caption{Graphical overview of matrix distribution on 4 ranks and computational operations on local matrix data.} \label{fig:mcholqrbgs}
\end{figure*}

The modified Cholesky-QR2 algorithm introduces a modification of CholeskyQR2 with Gram-Schmidt (mCQR2GS) by rearranging the order of computational tasks to reduce the larger number of panels required in CQRGS. The computational complexity and communication are therefore equivalent to the CQRGS algorithm with the same number of panels. This idea provides an adaptive paneling strategy that is tailored to the inherent properties of each matrix and optimizes the factorization process for both ill- and well-conditioned matrices. An example is shown in Fig.~\ref{fig:mcholqrbgs}, where the matrix is divided into 3 panels. 

\begin{algorithm}[ht]
  \caption{Modified Cholesky QR with Gram-Schmidt (mCQRGS)}\label{alg:mcholqrgs}
    \begin{algorithmic}[1]
      \Require{$A \in \mathbb{R}^{m\times n}$, number of panels $k$}
      \Ensure{$ Q \in \mathbb{R}^{m\times n} $ orthogonal and $ R \in \mathbb{R}^{n\times n} $ upper triangular matrix }
      \State $Q_1, R_{1,1} = CQR2(A_{:,1})$ \Comment{Orthogonalize first panel}
      \For {$j = 2\ldots k$}
        \State $Y := Q^T_{j-1}\ A_{:,j:k}$   \Comment{Projections of orthogonal panels on non-orthogonal}
        \State $A_{:,j:k} := A_{:,j:k} - Q_{j-1}\ Y$ \Comment{Update panels $A_{j},\ldots A_k$}
        \State $[R_{j-1,j}, R_{j-1,j+1},\ldots,R_{j-1,k}] := Y$
        \State $CQR(A_{:,j})$
        \State $A_{:,j} := A_{:,j} - Q_{1:j-1}\ Q^T_{1:j-1}\ A_{:,j} $ \Comment{Reorthogonalize current panel}
        \State $Q_j = CQR(A_{:,j})$
    \EndFor
  \end{algorithmic}
\end{algorithm}

The input matrix $A$ is partitioned and distributed among $P$ processors in the same manner as in CQRGS algorithm (see Fig.~\ref{fig:distA}) with partitioned of local block matrix $A_p$ into panels (Fig.~\ref{fig:mcholqrbgs}). The pseudocode for the modified algorithm is listed in Algorithm \ref{alg:mcholqrgs}. Each rank starts with orthogonalizing the first panel by applying full  CQR2 (line 1) and then reorthogonalize the trailing panels via the Gram-Schmidt process (line 3 i 4) in parallel. The algorithm then moves by computing the QR of the second panel using  CQR (line 6). However, the orthogonalization of this second panel is only partially completed at this stage. Therefore, another step of reorthogonalization with respect to the already orthogonal panels is required (line 7), followed by a call to CQR to compute fully orthogonal $Q_j$ (line 8). This careful process ensures the complete orthogonalization of the second panel and establishes its orthogonality to the first panel. The algorithm then proceeds to the third panel with the same routine as for the second panel.

\begin{figure}[h]
\centering
\includegraphics[width=0.4\textwidth]{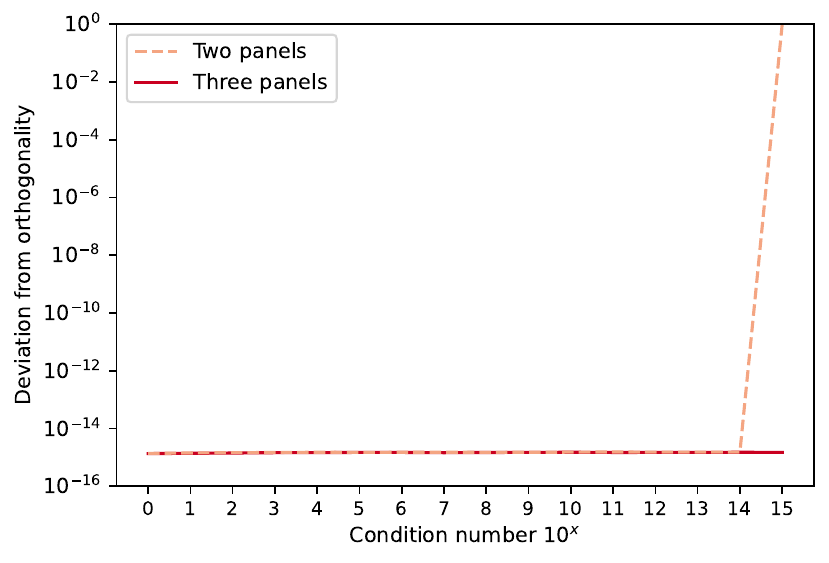}
\caption{Orthogonality of the $Q$ factor of mCQR2GS with 2 and 3 panels w.r.t. to the condition number on one node and 4 GPUs. Matrix size $30k \times 3k$.}
\label{fig:mcholqrbg_ortho}
\end{figure}

\begin{figure}[h]
\centering
\includegraphics[width=0.4\textwidth]{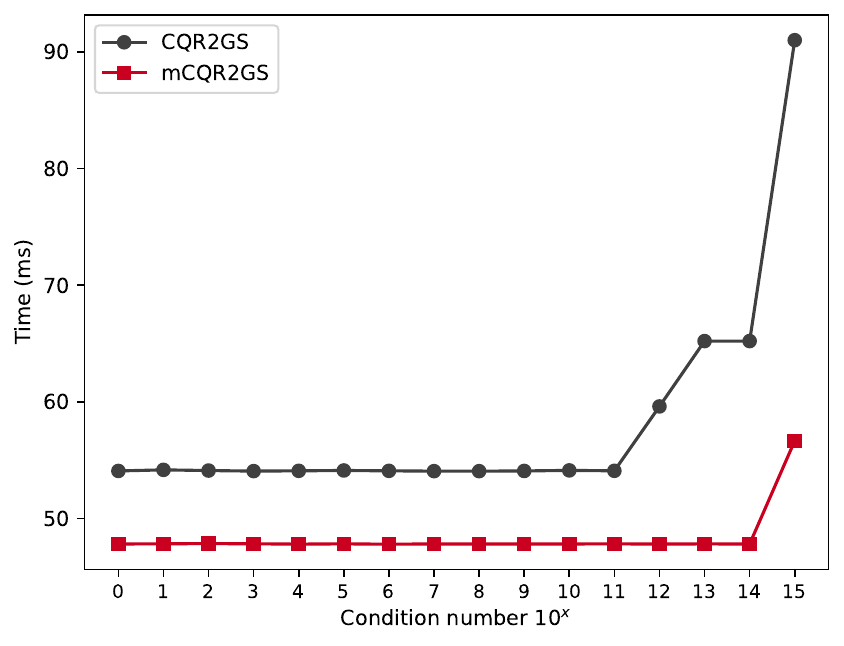}
\caption{Total execution time of CQR2GS and mCQR2GS when using the optimal number of panels on one node and 4 GPUs. Matrix size $30k \times 3k$.} 
\label{fig:mcholqrbg_time}
\end{figure}

Our analysis showed that the 3-panel strategy is optimal for our artificially generated use cases as illustrated in Fig.~\ref{fig:mcholqrbg_ortho}. As expected, the mCQR2GS algorithm still breaks down on our use-cases when the 2-panel strategy is used on matrices with a very high condition number ($\ge 10^{15}$). The reason is that the condition number of the first panel, according to the Eq.~\ref{eq:cond_bounds}, is upper bounded by the condition of the input matrix ($10^{15}$) and lower-bounded by half of the condition number ($\approx10^8$) which, in the best scenario of reaching the lower bound, results in a Gram matrix with condition number greater than $10^{15}$ and not fully semi-positive definite. Furthermore, the mCQR2GS requires less flops compared to CQR2GS as illustrated in Fig.~\ref{fig:mcholqrbg_time}, as it does not require the explicit construction of factor $R$ at the end (see Alg.~\ref{alg:cholqr2} step 3). Note that for up to the condition number $10^{11}$ both algorithms use 2-panel strategy as the optimal one. With the increasing condition number CQR2GS requires a larger number of panels to secure the stability and orthogonality, while mCQR2GS still requires only 2 panels, except for the last use-case in which 3 panels are needed to improve the orthogonality. 

\section{Scalability analysis}
\label{sec:scalability}

In the last part, we analyse the strong and weak scaling performance of the Modified CholeskyQR2 with Gram-Schmidt (mCQR2QR) and compare it with ScaLAPACK.
In our test cases, the 3-panel strategy for mCQR2GS is used for extremely ill-conditioned matrices. All tests achieve the required numerical stability and orthogonality close to the machine precision $u$. In the case of the GPU version, the NCCL communicator was used for collective communication instead of the CUDA-aware MPI, as NCCL achieves much better performance by significantly reducing the communication overhead.

\begin{figure}[]
\centering
\includegraphics[width=0.5\textwidth]{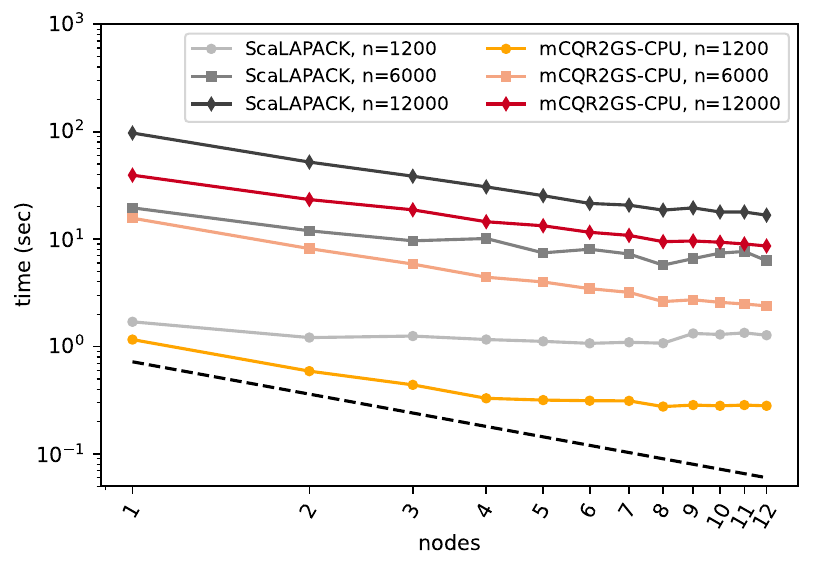}
\caption{Strong scaling of mCQR2GS-CPU and ScaLAPACK with artificial matrices $m=120k$, $n=\{1.2k, 6k,12k\}$ w.r.t the number of nodes. The dashed line is the ideal scaling.}
\label{fig:TTSallalg}
\end{figure}

\begin{figure}[]
\centering
\includegraphics[width=0.5\textwidth]{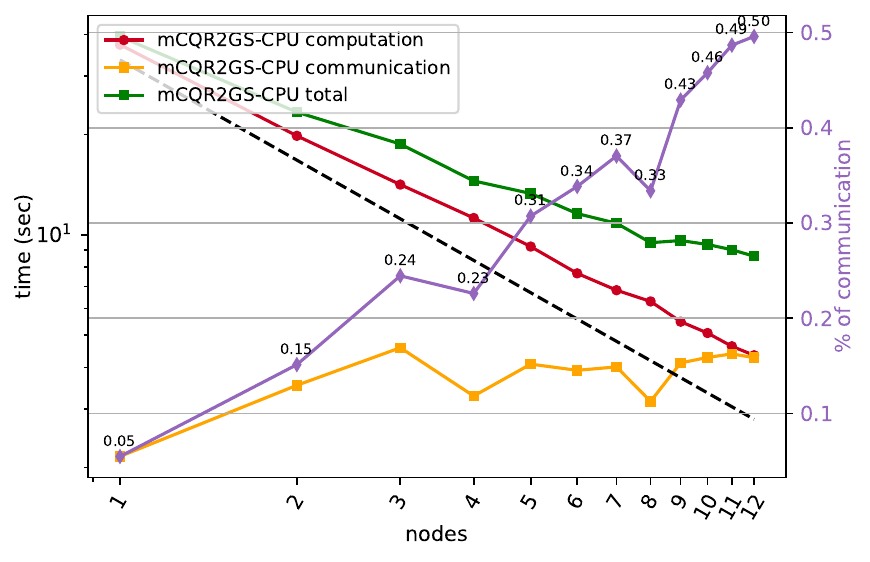}

\caption{Total execution time (green), execution times for calculation (red) and communication (yellow) and share of communication in the total execution time (purple) of mCQR2GS. Matrix size $120k \times 12k$.}
\label{fig:strongScaling-speedup}
\end{figure}

The strong scaling was tested on the 3 artificial matrices (see subsection~\ref{subsec:testMatrixSuite}) with the condition number $\kappa(10^4)$ and the dimensions $120k \times 1.2k$, $120k \times 6k$ and $120k\times 12k$ (the panels are $400, 2000$ and $4000$ wide respectively). Fig.~\ref{fig:TTSallalg} illustrates the strong scaling behaviour of the CPU-only version of mCQR2GS and ScaLAPACK. Note that the scalability of the mCQR2GS algorithm decreases with the number of nodes. The reason for this is the communication (operation \textit{Allreduce}), which is performed when building the Gram matrix (Algorithm \ref{alg:cholqrgs_distributed}, line 3) and does not scale with the number of nodes. In strong scaling tests, the width of the panel is fixed, which leads to a constant load in Gram reduction operations, while the communication load increases with the number of nodes, as shown in Fig.~\ref{fig:strongScaling-speedup} (yellow line). Although the computation parts scale close to the ideal line, the communication time remains constant or slightly increases, resulting in an overall lower scalability of the code as communication becomes more and more dominant, up to $50\%$ of the total execution time (purple line). The observed peaks in communication performance (Fig.~\ref{fig:strongScaling-speedup}, yellow line) on 4 and 8 nodes (16 and 32 MPI processes respectively) are the result of the internal NCCL optimisation of allreduce operations implemented as a binary tree.
For all tested matrices, the mCQR2GS CPU outperforms ScaLAPACK by up to $4.7\times$. 

Weak scaling experiments show the potential of the novel algorithm to perform computations on large matrices while keeping the load per process or node constant.
The tests were performed on both CPU and GPU partitions of the Supek supercomputer.
The ScaLAPACK configuration is set to 16 tasks (MPIs) per node and 8 threads per task, with the block height set to the number of rows divided by the number of MPI processes and the block width set to 32. This configuration achieved the best performance in our sweet spot analysis.

\begin{figure}[]
\centering
\includegraphics[width=0.5\textwidth]{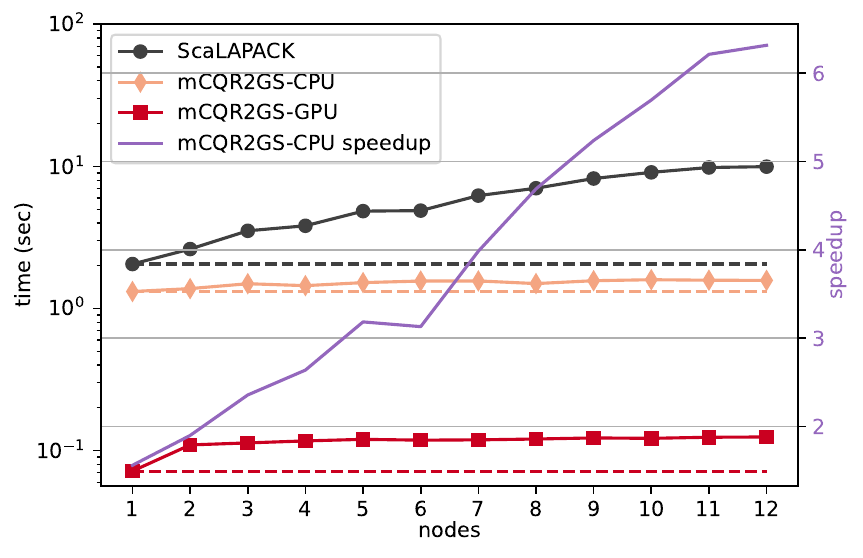}
\caption{Weak scaling of mCQR2GS-GPU, mCQR2GS-CPU and SCALAPACK with fixed block size per node $40k \times 3k$ and 4 MPI processes per node. Dashed lines are the ideal scaling.} 
\label{fig:TTWallalg}
\end{figure}

Fig.~\ref{fig:TTWallalg} shows that the weak scaling is nearly optimal for both the CPU and GPU versions of mCQR2GS, with the total execution time ranging from $71.78$ msec on one node to $124$ msec on 12 nodes for the GPU version. A steep time jump from one to two nodes ($1.53 \times$) is the NCCL communication overhead, which was not observed when all NCCL processes are on the same node. Although the introduction of inter-node communication is significant when switching to two nodes, it remains almost constant when we introduce more nodes. Note that the performance of ScaLAPACK is decreasing as the number of nodes increases. Since the number of columns is fixed, the matrices become thinner and thinner as the number of nodes increases (i.e.~fixed number of rows per process/node), resulting in a significant performance degradation for ScaLAPACK, which is tailored for general and square matrices instead of tall and-skinny matrices. Although the communication in mCQR2GS is significant compared to the total execution time, both the GPU and CPU versions of the code achieve a significant speedup compared to ScaLAPACK, with the CPU variant gaining more than $6\times$ and the GPU variant $80\times$ in speedup.

\section{Conclusion}
\label{sec:conclusion}
In this paper, we have discussed the advantages and competitiveness of several introduced algorithms, namely CQR2, sCQR3, CQR2GS and Householder QR, in terms of key aspects including: suitability to perform in distributed environments, numerical stability and the handling of tall-and-skinny matrices. Each of the above algorithms exhibits strengths in one or more of these areas. CQR2, for example, shows robust distributed parallelisation capabilities, but lack in numerical stability. The sCQR3 algorithm has good numerical stability, but with a tradeoff in terms of extra flops. On the other hand, while the Householder-based QR provides good numerical stability, it performs sub-optimally on tall and skinny matrices. 
We present a novel algorithm, which we call modified CholeskyQR2 with Gram-Schmidt (mCQR2GS), to compute the QR factorization of tall and skinny matrices on distributed multi-GPU architectures. Our approach attempts to find a balance in terms of all the above-mentioned features.

The novelty of the algorithm lies in complete orthogonalisation of the working panel before performing the Gram-Schmidt step, which is not the case in the original CholeskyQR2 with Gram-Schmidt. 
The novel algorithm achieves much better numerical stability compared to other solutions tailored to tall-and-skinny matrices, such as CholeskyQR2, especially in the case of extremely ill-conditioned matrices (condition number of up to $10^{16}$). Our algorithm proved to be faster than CQR2GS and outperforms ScaLAPACK on both distributed CPU and GPU systems by $6\times$ and $80\times$, respectively.

The most important tuning parameter for performance and stability is the panel width. By slicing the input matrix into smaller panels, the condition number of each panel decreases. However, there is a tradeoff in terms of performance when doing so. It is essential to tune this parameter to balance between performance and stability and for our testing architecture and matrix dataset (with equally distributed singular values), the optimal number of panels was 3. 
However, in the case of clustered singular values, our approach can not improve the loss of orthogonality or numerical stability. To address this issue, we plan to extend our solution with a shifting strategy to ensure stability in such extreme use-cases.

The main bottleneck of our algorithm is the collective communication, whose ratio to the total execution time increases by up to $50\%$ with an increasing number of processes. An ongoing effort is the implementation of a look-ahead approach to overlap the update of panels with computing the CholeskyQR of the next panel (Algorithm \ref{alg:mcholqrgs} lines 4 and 6).  
The communication in CholeskyQR-based algorithms is expensive compared to the computation, especially if CholeskyQR has to be repeated multiple times to improve the orthogonality of the factor $Q$, and increases with the number of processors. Moreover, the condition number steeply decreases as we proceed with the panel processing, opening up a space for further optimisation in reducing the number of flops by applying a runtime decision on how many repetitions of CholeskyQR to perform. 

The source code of the Shifted CholeskyQR3, CQR2GS and mCQR2GS versions is available on the GitHub pages~\footnote{https://github.com/HybridScale/CholeskyQR2-IM} and Zenodo~\cite{CholQR2GS-IM}.

\ifCLASSOPTIONcompsoc
  % The Computer Society usually uses the plural form
  \section*{Acknowledgments}
\else
  % regular IEEE prefers the singular form
  \section*{Acknowledgment}
\fi

This research was supported by the Croatian Science Foundation under grant number HRZZ-UIP-2020-02-4559 and the European Regional Development Fund under grant KK.01.1.1.01.009 - DATACROSS. All the computations were performed using the Advanced computing service provided by the University of Zagreb University Computing Centre - SRCE.

% Can use something like this to put references on a page
% by themselves when using endfloat and the captionsoff option.
\ifCLASSOPTIONcaptionsoff
  \newpage
\fi

% trigger a \newpage just before the given reference
% number - used to balance the columns on the last page
% adjust value as needed - may need to be readjusted if
% the document is modified later
%\IEEEtriggeratref{8}
% The "triggered" command can be changed if desired:
%\IEEEtriggercmd{\enlargethispage{-5in}}

% references section

% can use a bibliography generated by BibTeX as a .bbl file
% BibTeX documentation can be easily obtained at:
% http://mirror.ctan.org/biblio/bibtex/contrib/doc/
% The IEEEtran BibTeX style support page is at:
% http://www.michaelshell.org/tex/ieeetran/bibtex/
\bibliographystyle{IEEEtran}
\bibliography{references}
% argument is your BibTeX string definitions and bibliography database(s)
%\bibliography{IEEEabrv,../bib/paper}
%
% <OR> manually copy in the resultant .bbl file
% set second argument of \begin to the number of references
% (used to reserve space for the reference number labels box)
%\begin{thebibliography}{1}

%\bibitem{IEEEhowto:kopka}
%H.~Kopka and P.~W. Daly, \emph{A Guide to \LaTeX}, 3rd~ed.\hskip 1em plus
%  0.5em minus 0.4em\relax Harlow, England: Addison-Wesley, 1999.

%\end{thebibliography}

% biography section
% 
% If you have an EPS/PDF photo (graphicx package needed) extra braces are
% needed around the contents of the optional argument to biography to prevent
% the LaTeX parser from getting confused when it sees the complicated
% \includegraphics command within an optional argument. (You could create
% your own custom macro containing the \includegraphics command to make things
% simpler here.)
%\begin{IEEEbiography}[{\includegraphics[width=1in,height=1.25in,clip,keepaspectratio]{mshell}}]{Michael Shell}
% or if you just want to reserve a space for a photo:

%\begin{IEEEbiography}{Michael Shell}
%Biography text here.
%\end{IEEEbiography}

% if you will not have a photo at all:
\vspace{-1cm}

\begin{IEEEbiographynophoto}{Nenad Miji\' c}
received his M.S. degree in Physics from the University of Zagreb in 2017. He is currently working on his PhD under the supervision of Davor Davidović. His research interests include code portability and eigenvalue solvers on heterogeneous distributed systems.
\end{IEEEbiographynophoto}
% insert where needed to balance the two columns on the last page with
% biographies
%\newpage

\vspace{-1cm}

\begin{IEEEbiographynophoto}{Abhiram Kaushik}
is a postdoctoral researcher at the Ruđer Bošković Institute. He received his PhD in Physics from the Indian Institute of Science in 2019. His research interests include the application of high-performance computing techniques to problems in high-energy physics and numerical linear algebra. He has published 11 papers in peer-reviewed international scientific journals.
\end{IEEEbiographynophoto}

\vspace{-1cm}

\begin{IEEEbiographynophoto}{Davor Davidovi\' c}
is a senior research associate at the Ruđer Bošković Institute. He received his PhD in computer science from the University of Zagreb in 2014. His research interests are parallel and distributed computing, code optimisation, hybrid and GPU programming, and scalable algorithms for linear algebra. He has participated in more than 15 international research projects (FP7, H2020, COST). He has published 34 original scientific papers in international journals and in conference proceedings.
\end{IEEEbiographynophoto}

% You can push biographies down or up by placing
% a \vfill before or after them. The appropriate
% use of \vfill depends on what kind of text is
% on the last page and whether or not the columns
% are being equalized.

%\vfill

% Can be used to pull up biographies so that the bottom of the last one
% is flush with the other column.
%\enlargethispage{-5in}

% that's all folks
\end{document}